\documentclass[twocolumn,showpacs,aps,amsmath,amssymb,floatfix,superscriptaddress]{revtex4}
\pdfoutput=1 
\usepackage{graphicx,eucal}

\begin{document}
\title{Symmetric Exclusion Process with a Localized Source}
\author{P. L. Krapivsky}
\affiliation{Department of Physics, Boston University, Boston, Massachusetts 02215, USA}

\begin{abstract}
We investigate the growth of the total number of particles in a symmetric exclusion process driven by a localized source. The average total number of particles entering an initially empty system grows with time as $\sqrt{t}$ in one dimension, $t/\ln t$ in two dimensions, and linearly in higher dimensions. In one and two dimensions, the leading asymptotic behaviors for the average total number of particles are independent of the intensity of the source.  We also discuss fluctuations of the total number of particles and determine the asymptotic growth of the variance in one dimension.
\end{abstract}

\maketitle

\section{Introduction}

A system of particles undergoing symmetric random walks on a lattice under the constraint that multiple occupancy is forbidden, a so-called symmetric exclusion process (SEP), has been extensively investigated, especially in one spatial dimension  (see, e.g., books and reviews \cite{Spohn,KL99,L99,SZ95,D98,S00,BE07}). A surprising feature of the SEP, at least at first sight, is that the average density evolves as if there were no exclusion. (More subtle characteristics are affected by exclusion, e.g., the mean-square displacement of the tagged particle exhibits a remarkable sub-diffusive growth, $\langle x^2\rangle \sim \sqrt{t}$, in one dimension \cite{Harris_65,Levitt_73,PMR_77,AP_78,book}; in higher dimensions, the diffusion behavior of the tagged particle is restored and only the amplitude of the mean-square displacement acquires the dependence on the concentration.) This may lead to the impression that the exclusion property is irrelevant as long as the average characteristics are concerned. Further, the peculiarities of the SEP arise in one dimension; when $d>1$, the SEP and a system of non-interacting random walks exhibit similar qualitative behaviors. 
 
The above conclusions hold in the absence of sources. If particles enter the system through external reservoirs, the exclusion property may become crucial. Here we analyze the SEP with a localized source and find that exclusion plays an important role, particularly in low dimensions $d\leq 2$. More precisely, we show that the critical dimension is $d_c=2$, viz. similar behaviors arise when $d\geq 3$. (When $d=2$, the difference with the higher-dimensional behavior is logarithmic, i.e., rather small.)

We study the SEP on the $d-$dimensional hyper-cubic lattice  $\mathbb{Z}^d$. We set to $1/d$ the hopping rate to each of the $2d$ neighboring sites, so that the total hopping rate is equal to 2 for all $d$. (The hopping event is allowed only when the destination site is empty.)  We denote by $F$ the flux of particles  to the origin, that is, the rate at which new particles would be arriving if the origin were always empty. We want to understand how the total number of particles $N(t)$ grows.  (The source is turned on at $t=0$; we assume that the system is initially empty.) 

The quantity $N(t)$ is a random variable. In the large time limit, $N(t)$ is concentrated near its average, that is, fluctuations are relatively small. We show that the average $\langle N\rangle$ grows as 
\begin{equation}
\label{Nav}
 \langle N\rangle \simeq 
\begin{cases}
4\sqrt{t/\pi}                    &d=1\\
2\pi\, t/\ln t                     &d=2\\
\Phi_d(F)\,t                   &d>2
\end{cases}
\end{equation}
in the long time limit. Intriguingly, the leading asymptotic behaviors in one and two dimensions are independent of the flux $F$ as long as it is positive. Also, in one and two dimensions the total number of particles entering the system is a negligible fraction of the number of particles, $Ft$ on average, which would have entered in the no-exclusion case. In three and higher dimensions, the average total number of particles entering the system grows linearly with time. The renormalized flux $\Phi_d(F)$ admits a simple expression via the `bare' flux $F$ and a Watson integral $W_d$ [see \eqref{Watson}]
\begin{equation}
\label{JF_3}
\Phi_d(F) = \frac{F}{1+FW_d}
\end{equation}
Thus the renormalized flux grows linearly with bare $F$ when $F\ll 1$, and saturates to $W_d^{-1}$ in the $F\to\infty$ limit; needless to say, $\Phi_d(F)<F$. 

We also discuss fluctuations of the total number of injected particles. Specifically, we argue that the variance $\langle N^2\rangle_c\equiv \langle N^2\rangle - \langle N\rangle^2$ exhibits an asymptotic growth 
\begin{equation}
\label{N_variance}
 \langle N^2\rangle_c \simeq 
\begin{cases}
V_1\sqrt{t}                     &d=1\\
V_2\,\tfrac{t}{\ln t}          &d=2\\
\Phi_d(F)\,t                    &d>2
\end{cases}
\end{equation}
and we determine the amplitude $V_1$ in one dimension 
\begin{equation}
\label{var_1d}
V_1 = \frac{4 \big(3-2\sqrt{2}\big)}{\sqrt{\pi}}
\end{equation}
The computation leading to this result is performed in the $F=\infty$ limit, but we argue that the amplitude is independent on the flux $F$. 

We study the SEP with a localized source, but the results are more widely applicable. For instance, some 
stochastic processes are mathematically similar, or even isomorphic, to the SEP with a localized source with infinitely strong flux. Examples include monomer-monomer catalysis and the voter model \cite{PK,LF,Mauro}. Both models are particularly natural in two dimensions and they have been analyzed for $d\leq 2$; some of the results of Refs.~\cite{PK,LF,Mauro} are equivalent to the predictions \eqref{Nav} in one and two dimensions. The spreading of very thin wetting films has been described by a SEP-like model (see \cite{Gleb_rev} for a review). The natural dimensionality of the substrate is $d=2$ and the average injected mass has been computed \cite{Gleb} for $d\leq 2$ , and has been shown to compare favorably with experimental observations. More complicated SEP-like models with a localized source have also been studied. One recent example \cite{darko} involves synthetic molecular motors, so-called molecular spiders \cite{spiders}; in contrast to particles in the SEP,  the underlying model of molecular spiders \cite{memo} is non-Markovian as the motion of spiders affect the substrate.

The rest of this paper is organized as follows. A heuristic explanation of asymptotic behaviors \eqref{Nav} is given in Sec.~\ref{heuristic}. In Sec.~\ref{inf} we present an exact solution in the limiting case of the infinitely strong source, from which we determine exact asymptotic behaviors. The general case of finite flux is treated in Sec.~\ref{gen} where we derive the renormalized flux, Eq.~\eqref{JF_3}, when $d>2$, and we compute sub-leading corrections in one dimension. In Sec.~\ref{fluct} we discuss fluctuations. In particular, we analytically determine the variance in the one-dimensional setting. Section~\ref{summary} contains a summary. 

\section{Heuristic Derivation}
\label{heuristic}

Here we give a heuristic explanation of the growth laws \eqref{Nav}. We consider the simplest situation when the source is infinitely strong, $F=\infty$. In this case, the density at the origin is $\rho_{\bf 0}=1$ for all $t\geq 0$. The initial density in other sites is zero. In one dimension, the density decays from 1 near the origin to zero on distances exceeding the diffusive scale $\sqrt{t}$. The density profile is roughly linear in the diffusive layer:
\begin{equation}
\label{density_approx_1}
 \rho \approx 
\begin{cases}
1-|x|/\sqrt{t}                 & |x|<\sqrt{t}\\
0                                 & |x|>\sqrt{t}
\end{cases}
\end{equation}
Therefore, $\langle N\rangle \sim\sqrt{t}$, in agreement with \eqref{Nav}. 

In two dimensions, the density is also time-dependent. Using the well-known asymptotic (see, e.g., \cite{fpp})
\begin{equation}
\label{density_approx_2}
\rho\simeq 1 - \frac{\ln r}{\ln\sqrt{t}}
\end{equation}
we obtain
\begin{equation*}
\langle N\rangle \sim \int_1^{\sqrt{t}}dr\,r\,\left[1 - \frac{\ln r}{\ln\sqrt{t}}\right]
\sim \frac{t}{\ln t} 
\end{equation*}

In three dimensions, the density becomes stationary. Thus one must solve the Laplace equation $\nabla^2 \rho=0$. An appropriate solution is $\rho(r)\sim r^{-1}$ (Coulomb's potential), or more generally $\rho(r)\sim r^{-(d-2)}$ when $d>2$. There are no particles on distances far exceeding $\sqrt{t}$. The average number of particles is estimated by integrating over the ball of radius of the order of $\sqrt{t}$. This gives 
\begin{equation}
\label{N-estimate}
\langle N\rangle \sim \int_1^{\sqrt{t}}dr\,r^{d-1}\,\rho(r)\sim t
\end{equation}
where in the last step we used $\rho(r)\sim r^{-(d-2)}$. The linear growth law \eqref{N-estimate} applies when $d>2$. 

\section{Infinitely Strong Source}
\label{inf}

The case of an infinitely strong source is exactly solvable as far as the average quantities are concerned. The average density $\rho_{\bf x}(t)$ satisfies the diffusion equation
\begin{equation}
\label{diff-eq}
\frac{\partial \rho}{\partial t}= \frac{1}{d}\,\nabla^2 \rho
\end{equation}
Here ${\bf x} = (x_1,\ldots, x_d)\in \mathbb{Z}^d$ is the lattice site and $\nabla^2$ denotes the discrete Laplace operator; e.g.,  in one dimension $\nabla^2 \rho_x = \rho_{x+1}-2\rho_x+\rho_{x-1}$.  The entire hopping rate is set to 2, so the individual hopping rates are equal to $1/d$ as any site of the hyper-cubic lattice has $2d$ nearest neighbors. The initial condition reads
\begin{equation}
\label{diff-in}
\rho_{\bf x}(t=0)=\delta_{{\bf x},{\bf 0}}
\end{equation}
An infinitely strong flux is represented by the boundary condition 
\begin{equation}
\label{diff-BC}
\rho_{\bf 0}(t)=1
\end{equation}
assuring that there is always a particle at the origin. Equations \eqref{diff-eq}--\eqref{diff-BC} admit an exact solution, from which we will extract leading asymptotic behaviors of the average total number of particles:
\begin{equation}
\label{exact-inf}
 \langle N\rangle \simeq 
\begin{cases}
4\sqrt{t/\pi}                    &d=1,\\
2\pi\, t/\ln t                     &d=2,\\
t/W_d                            &d>2.
\end{cases}
\end{equation}
Here $W_d$ are the so-called Watson's integrals \cite{Watson}
\begin{equation}
\label{Watson}
W_d=\int_0^{2\pi}\ldots\int_0^{2\pi} \frac{1}{Q({\bf q})}\prod_{i=1}^d \frac{dq_i}{2\pi}
\end{equation} 
where ${\bf q} = (q_1,\ldots,q_d)$ and 
\begin{equation}
\label{Qq}
Q({\bf q}) = \frac{2}{d}\sum_{i=1}^d (1-\cos q_i)
\end{equation}

Watson's integrals often appear in problems involving lattice Laplacians. 
In three dimensions, the Watson integral \eqref{Watson} has been expressed \cite{GZ_77} via Euler's gamma functions:
\begin{eqnarray*}
W_3 &=& \frac{\sqrt{6}}{64\,\pi^3}\, 
\Gamma\left(\frac{1}{24}\right)\, \Gamma\left(\frac{5}{24}\right)\, 
\Gamma\left(\frac{7}{24}\right)\, \Gamma\left(\frac{11}{24}\right)\\
&=& 0.75819303\ldots
\end{eqnarray*} 

The solution to \eqref{diff-eq}--\eqref{diff-BC} can be established using various approaches. Here we outline a derivation following an approach of Refs.~\cite{PK,LF,Mauro,book} which holds in arbitrary spatial dimension and gives exact results for Laplace transforms of the basic quantities. Asymptotic behaviors of the Laplace transforms imply asymptotic large time behaviors. We then use another method and obtain an explicit solution in one dimension in a more direct way, namely, without using the Laplace transform. 

\subsection{Solution via Laplace Transform}

A solution to Eqs.~\eqref{diff-eq}--\eqref{diff-in}, i.e., the solution of the problem without the boundary condition \eqref{diff-BC}, is merely the lattice Green function $I_{\bf x}(2t/d)e^{-2t}$. Here  
\begin{equation}
\label{multi_B}
I_{\bf x}(\tau) = \prod_{j=1}^d I_{x_j}(\tau)
\end{equation}
is the shorthand notation for the product of the modified Bessel functions. To maintain the validity of the boundary condition $\rho_{\bf 0}(t)=1$ throughout the evolution, we employ a simple trick: We add particles to the origin at a certain rate $\Phi_d(t)$ which we choose to fulfill the boundary condition. Since the governing equation is linear, the general solution to \eqref{diff-eq}--\eqref{diff-in} with a source is a linear combination
\begin{equation}
\label{source}
\rho_{\bf x}(t)=
I_{\bf x}(\tfrac{2t}{d})\,e^{-2t}+\int_0^t d\tau\, \Phi_d(t-\tau)\,I_{\bf x}(\tfrac{2\tau}{d})\,e^{-2\tau}
\end{equation}
The integral term is the contribution due to the source $\Phi_d(\tau)d\tau$ which is added at the origin during the time interval $(\tau, \tau + d\tau)$; the original source $\delta(t)$ yields the first term on the right-hand side of \eqref{source}. At the origin
\begin{equation}
\label{int-eq}
1 = \left[I_0(\tfrac{2t}{d})\right]^d e^{-2t}  +\! \int_0^t d\tau\,\Phi_d(t-\tau)\left[I_0(\tfrac{2\tau}{d})\right]^d e^{-2\tau}
 \end{equation}
This integral equation determines the strength $\Phi_d(t)$ of the source. To extract an explicit expression we notice the convolution structure of the integral in \eqref{int-eq}. Hence we apply the Laplace transform. This yields a neat relation
\begin{equation}
\label{IJ}
\widehat{\Phi}_d(s) = \frac{1}{s\widehat{B}_d(s)}-1
\end{equation}
between the Laplace transform of the strength of the source and the Laplace transform of the power of the Bessel function $I_0$:
\begin{subequations}
\begin{align}
  \label{Js}
&\widehat{\Phi}_d(s)   = \int_0^\infty dt\, e^{-s t}\,\Phi_d(t)\\
  \label{Ids}
&\widehat{B}_d(s) = \int_0^\infty dt\, e^{-s t}\left[I_0(\tfrac{2t}{d})\right]^d \,e^{-2t}
\end{align}
\end{subequations}
Using the integral representation of the modified Bessel function, $I_0(t)=\int_0^{2\pi}\frac{dq}{2\pi}\,\exp(t\cos q)$, and performing the integration over $t$, we re-write \eqref{Ids} as an integral
\begin{equation}
\label{Bd_s}
\widehat{B}_d(s)=\int_0^{2\pi}\cdots\int_0^{2\pi} \frac{1}{s+Q({\bf q})}\prod_{i=1}^d \frac{dq_i}{2\pi}
\end{equation} 
with $Q({\bf q})$ given by \eqref{Qq}. Expanding the right-hand side of \eqref{Bd_s} we determine the small $s$ behavior  
\begin{equation}
\label{Is-lead}
\widehat{B}_d(s)\simeq
\begin{cases}
2^{-1} s^{-1/2}            &d=1,\\
(2\pi)^{-1}\ln(1/s)        &d=2,\\
\widehat{B}_d(0)         &d>2,
\end{cases}
\end{equation}
where $\widehat{B}_d(0) = W_d$ for $d>2$. Combining Eqs.~\eqref{IJ} and \eqref{Is-lead} we find the leading $s\to 0$ behavior of the Laplace transform of the strength of the source 
\begin{equation}
\label{Js-lead}
\widehat{\Phi}_d(s)\simeq 
\begin{cases}
2\,s^{-1/2}                                   &d=1,\\
2\pi\, s^{-1}[\ln(1/s)]^{-1}             &d=2,\\
(W_d\, s)^{-1}                             &d>2,
\end{cases}
\end{equation}
Converting \eqref{Js-lead} we obtain the large time asymptotic of the strength of the source
\begin{equation}
\label{Jt-lead}
\Phi_d(t)\simeq 
\begin{cases}
2\,\pi^{-1/2}\,t^{-1/2}   &d=1,\\
2\pi/\ln t                      &d=2,\\
1/W_d                        &d>2.
\end{cases}
\end{equation}
By inserting \eqref{Jt-lead} into the relation $\langle N(t)\rangle = \int_0^t d\tau\, \Phi_d(\tau)$ we arrive at the announced asymptotic \eqref{exact-inf}. 

Using Eqs.~\eqref{source} and \eqref{Jt-lead} we can additionally extract the asymptotic behavior of the density. In one dimension, it is actually simpler to employ a continuum approach from the very beginning. Thus we must solve the initial-boundary value problem
\begin{equation*}
\partial_t \rho = \partial_{xx} \rho, \quad \rho(x=0,t)=1, \quad \rho(x, t=0)=0
\end{equation*}
for the average density $\rho(x,t)$. The solution is 
\begin{equation}
\label{density_1d}
\rho(x,t) = \text{erfc}\left(\frac{|x|}{\sqrt{4t}}\right)
\end{equation}
where $\text{erfc}(u)\equiv \frac{2}{\sqrt{\pi}}\int_u^\infty dv\,e^{-v^2} = 1 - \text{erf}(u)$ is an error function. The asymptotically exact result \eqref{density_1d} is the corrected version of \eqref{density_approx_1}.

In deducing the asymptotic behavior of the density in two dimensions, we first recall that
\begin{equation}
\label{Bessel_Gauss}
e^{-t}I_n(t)\simeq \frac{1}{\sqrt{2\pi t}}\,e^{-n^2/2t}
\end{equation}
in the scaling region
\begin{equation}
t\to\infty, \quad n\to\infty, \quad \frac{n}{\sqrt{t}} = \text{finite}
\end{equation}
We now simplify the dominant integral term in Eq.~\eqref{source} by using \eqref{Jt-lead} and \eqref{Bessel_Gauss} to yield
\begin{equation}
\label{density_2d}
\rho_{\bf x}(t) = (\ln t)^{-1}\,E_1\!\big(\tfrac{r^2}{2t}\big)
\end{equation}
Here $r^2\equiv x_1^2+x_2^2$ and $E_1(z)=\int_1^\infty \frac{du}{u}\,e^{-zu}$ is an exponential integral. Using the asymptotic behavior of the exponential integral, $E_1(z)=-\ln z +\mathcal{O}(1)$ as $z\to 0$ \cite{BO78}, we see that Eq.~\eqref{density_approx_1} provides the leading asymptotic not too far from the source, namely when $r\ll \sqrt{t}$. 

In three and higher dimensions it suffices to consider the final (stationary) density. Therefore we put $t=\infty$ into Eq.~\eqref{source} and find 
\begin{equation}
\label{density_3d}
\rho_{\bf x} = \frac{1}{W_d} \int_0^{2\pi}\cdots\int_0^{2\pi} \frac{1}{Q({\bf q})} 
\prod_{i=1}^d \frac{\cos(q_i x_i)\, dq_i}{2\pi}
\end{equation} 
In deriving Eq.~\eqref{density_3d} from Eq.~\eqref{source} we have also used the integral representation \cite{BO78}
\begin{equation*}
I_n(\tau)=\int_0^{2\pi}\frac{dq}{2\pi}\,e^{\tau \cos q}\cos(nq)
\end{equation*}
of the Bessel function. For $d>2$, the steady state solution \eqref{density_3d} is clearly non-trivial. For instance, at the sites closest to the origin the density is $\rho'=1-(2W_d)^{-1}$; in three dimensions, $\rho'\approx 0.34053733$. 

\subsection{Explicit Solution in One Dimension}

Due to the $\rho_{-j}=\rho_j$ symmetry, it suffices to consider the densities $\rho_j$ to the right of the origin, $j>0$. Thus we must solve
\begin{equation}
\label{diff-1d}
\dot \rho_j = \rho_{j+1} - 2\rho_j + \rho_{j-1}
\end{equation}
subject to the initial condition $\rho_j(t=0)=0$ and the boundary condition \eqref{diff-BC}. 
We can instead consider \eqref{diff-1d} on the entire lattice and impose the initial condition
\begin{equation}
\label{IC:1d}
\rho_j(t=0) = 
\begin{cases}
2   &j<0\\
1   &j=0\\
0   &j>0
\end{cases}
\end{equation}
Then the boundary condition $\rho_0 = 1$ will manifestly hold. (The resulting solution is applicable, of course, only to $j\geq 0$; for $j<0$, proper densities are recovered from relation $\rho_{-j}=\rho_j$.) The solution of the discrete in space diffusion equation \eqref{diff-1d} subject to the initial condition \eqref{IC:1d} is straightforward:
\begin{equation}
\label{rj:1d}
\rho_j = e^{-2t}\left[I_j(2t)+2\sum_{k>j}I_k(2t)\right]
\end{equation}
The average total number of particles to the right of the origin is $\sum_{j>0}\rho_j$, and therefore
\begin{equation}
\label{Nav:1d}
\langle N\rangle = \rho_0+2\sum_{j>0}\rho_j
\end{equation}
Plugging \eqref{rj:1d} into \eqref{Nav:1d} and massaging the sums we get
\begin{equation}
\label{Nav:sum}
\langle N\rangle = e^{-2t}\left[I_0(2t)+4\sum_{k>0} k I_k(2t)\right]
\end{equation}
The sum on the right-hand side of Eq.~\eqref{Nav:sum} can be expressed through the Bessel functions $I_0$ and $I_1$ to give
\begin{equation}
\label{Nav:simple}
\langle N\rangle = e^{-2t}\left[I_0(2t)+4tI_0(2t)+4t I_1(2t)\right]
\end{equation}
The leading large time behavior of $\langle N(t)\rangle$ agrees with \eqref{exact-inf}, and as a byproduct of having a compact exact result \eqref{Nav:simple} we can also extract sub-leading corrections:
\begin{equation*}
\langle N\rangle = 4\sqrt{\frac{t}{\pi}}\left[1 + \frac{1}{16}\,t^{-1}+\frac{13}{2048}\,t^{-2}+\ldots\right]
\end{equation*}

\section{Finite Flux}
\label{gen}

Here we investigate the general case when the flux is finite, $F<\infty$. In one dimension, for instance, we need to solve Eq.~\eqref{diff-1d} away from the origin, while at the origin the average local density obeys
\begin{equation}
\label{1d-origin}
\dot \rho_0= 2(\rho_{1} - \rho_0) + F(1-\rho_0). 
\end{equation}
Due to the symmetry, $\rho_j = \rho_{-j}$, and it suffices again to consider only $j\geq 0$.

In one and two dimensions the leading asymptotic behavior is independent of $F$ for any $F>0$. For instance, in one dimension $\dot \rho_0\to 0$ and $(\rho_{1} - \rho_0)\to 0$ as $t\to\infty$, and therefore Eq.~\eqref{1d-origin} tells us that $\rho_0\to 1$. Thus the boundary condition \eqref{diff-BC} is asymptotically correct, and hence the leading asymptotic behaviors of all densities and of the total average number of particles are the same as in the case of the infinitely strong source. The two-dimensional case is  treated similarly. In three and higher dimensions, the average density at the origin exceeds the average densities at neighboring sites even as $t\to\infty$ and hence $\rho_{\bf 0} < 1$.

\subsection{Asymptotic Behavior in One Dimension}

To investigate the emergence of the asymptotic behavior in a quantitative manner we again employ the Laplace transform. Equations \eqref{diff-1d} and \eqref{1d-origin} become
\begin{equation}
\label{Lap_1d}
\begin{split}
s\widehat{\rho}_j &= \widehat{\rho}_{j+1} - 2\widehat{\rho}_j + \widehat{\rho}_{j-1}\,, \quad j>0 \\
s\widehat{\rho}_0 &= 2(\widehat{\rho}_1 - \widehat{\rho}_0) + F(s^{-1} - \widehat{\rho}_0)
\end{split}
\end{equation}
where
\begin{equation*}
\widehat{\rho}_j (s) = \int_0^\infty dt\,e^{-st} \rho_j(t)
\end{equation*}
Equations \eqref{Lap_1d} admit an exponential solution
\begin{equation}
\label{density_Aa}
\widehat{\rho}_j(s) = A(s) [a(s)]^j
\end{equation}
Plugging this ansatz into \eqref{Lap_1d} we get
\begin{equation}
\label{Aa}
A = \frac{1}{s}\,\frac{F}{F +\sqrt{s^2+4s}}, \quad
a = \frac{s+2-\sqrt{s^2+4s}}{2}
\end{equation}
The total average number of particle is
\begin{equation*}
\langle N(t)\rangle =\sum_{j=-\infty}^\infty \rho_j(t) = -\rho_0(t) + 2\sum_{j=0}^\infty \rho_j(t)
\end{equation*}
and therefore its Laplace transform is
\begin{equation}
\label{Lap_N}
\int_0^\infty dt\,e^{-st} \langle N(t)\rangle = A(s)\,\frac{1+a(s)}{1-a(s)}
\end{equation}
Using \eqref{Aa} we expand the right-hand side of \eqref{Lap_N} in the $s\to 0$ limit to yield
\begin{equation*}
A(s)\,\frac{1+a(s)}{1-a(s)} = 2s^{-3/2}-\frac{4}{F}\,s^{-1}+\left(\frac{8}{F^2}+\frac{1}{4}\right)s^{-1/2}+\ldots
\end{equation*}
Using this expansion in conjunction with \eqref{Lap_N} we deduce the large time expansion of the average total number of particles
\begin{equation}
\label{N_av_gen}
\langle N(t)\rangle = 4\sqrt{\frac{t}{\pi}} -\frac{4}{F} + \left(\frac{8}{F^2}+\frac{1}{4}\right)\frac{1}{\sqrt{\pi t}}+\ldots
\end{equation}
This asymptotic expansion confirms our assertion that the leading behavior is universal (that is, independent on $F$). 

We can also find the large time behavior of the densities to substantiate the claims $\rho_0\to 1$ and $\rho_1\to 1$ which have been previously made. Indeed, using \eqref{density_Aa} and \eqref{Aa} we establish the small $s$ expansions of the Laplace transforms
\begin{equation*}
\begin{split}
\widehat{\rho}_0 & = s^{-1} - \frac{2}{F}\,s^{-1/2} + \ldots \\
\widehat{\rho}_1 & = s^{-1} - \left(\frac{2}{F}+1\right)s^{-1/2} + \ldots
\end{split}
\end{equation*}
{}from which we deduce the $t\to\infty$ behaviors
\begin{equation}
\label{r_01}
\begin{split}
\rho_0 & = 1 - \frac{2}{F}\,\frac{1}{\sqrt{\pi t}} + \ldots \\
\rho_1 & = 1 - \left(\frac{2}{F}+1\right)\frac{1}{\sqrt{\pi t}}  + \ldots
\end{split}
\end{equation}

\subsection{Renormalized Flux in Three and Higher Dimensions}

In three and higher dimensions, the flux $F$ affects leading asymptotic behaviors. The  renormalized flux 
is time-independent in the long time limit, and therefore \eqref{source} gives 
\begin{equation}
\label{stationary}
\rho_{\bf x} = \Phi_d\int_0^\infty d\tau\,I_{\bf x}(\tfrac{2\tau}{d})\,e^{-2\tau}
\end{equation}
in the long time limit. Recalling \eqref{multi_B} we see that the density at the origin is 
\begin{equation}
\label{Jd_1}
\rho_{\bf 0} = \Phi_d\int_0^\infty d\tau\,\left[I_0(\tfrac{2\tau}{d})\right]^d\,e^{-2\tau} = \Phi_d W_d
\end{equation}
On the other hand, the renormalized flux is related with the density at the origin via
\begin{equation}
\label{Jd_2}
\Phi_d = F(1-\rho_{\bf 0})
\end{equation}
Combining \eqref{Jd_1} and \eqref{Jd_2} we arrive at the announced expression \eqref{JF_3} for the renormalized flux.

\section{Fluctuations}
\label{fluct}

The average numbers of injected particles [see Eq.~\eqref{exact-inf}] are lattice-independent when $d\leq 2$. This occurs because in one and two dimensions the spatial scale where the density varies is growing with time, so that the lattice structure is asymptotically irrelevant.  Hence the average quantities, e.g. the average densities [Eqs.~\eqref{density_1d} and \eqref{density_2d}], can be established in the realm of continuum approaches.  To probe fluctuations in low dimensions, $d\leq 2$, one can also utilize continuum approaches. 

For $d\geq 3$, the results are lattice-dependent.  The average density is stationary when $d\geq 3$ and this feature simplifies the problem. Namely, the total number of injected particles should be a Poisson distributed random quantity. In the long time limit, it becomes a Gaussian distribution with equal (in the leading order) average and variance $\langle N^2\rangle_c\equiv \langle N^2\rangle - \langle N\rangle^2$. In one and two dimensions, one still expects the average $\langle N\rangle$ and the variance $\langle N^2\rangle_c$ to exhibit the same dependence on time, but $\langle N\rangle \ne \langle N^2\rangle_c$ as there is no reason for the equality. 

These arguments lead to the announced results \eqref{N_variance}. The amplitudes $V_1$ and $V_2$ characterizing the variance in one and two dimensions should not depend on the flux $F$. The reason is the same as in the case of the average $\langle N\rangle$, although for the latter quantity the evidence is much stronger, and in one dimension we even have an exact solution (for the Laplace transform) and the asymptotic expansion \eqref{N_av_gen} which explicitly show that $F$ affects only sub-leading corrections. 

In this section we derive the amplitude in one dimension, Eq.~\eqref{var_1d}. In the computation we set $F=\infty$;  as we stated above, the result \eqref{var_1d} is apparently universal, namely it holds for any $F>0$. 

The variance can be computed by various techniques. In one dimension, one can use exact methods. To determine the most interesting long time behavior, it is easier to employ continuum approaches. These methods are, in principle, applicable both in one dimension and two dimensions, but the former case is much more tractable, so we consider only a one-dimensional setting. We shall compute the variance using fluctuating hydrodynamics (see \cite{Spohn,KL99,KM2011}). According to this approach one should solve a Langevin equation \cite{Spohn,KL99}
\begin{equation}
\label{Langevin}
     \partial_t q = \partial_{xx} q+\partial_x \left[\sqrt{\sigma(q)} \,\xi(x,t)\right]
\end{equation}
for the fluctuating particle density $q(x,t)$. Here $\xi(x,t)$ is a Gaussian white noise with standard correlations 
\begin{equation}
\left\langle \xi(x,t)\right\rangle=0, \quad 
\left\langle \xi(x,t)\xi(x^{\prime},t^{\prime})\right\rangle=\delta(x-x^{\prime})\, \delta(t-t^{\prime})
\label{noise}
\end{equation} 
The quantity $\sigma(q)$ characterizing fluctuations of the current is known \cite{Spohn,KL99} to be $\sigma(q) = 2q(1-q)$ for the SEP. 

When $F=\infty$, there is no interaction (for all $t>0$) between regions to the left and to the right of the source. Consider the $x>0$ region. Linearizing the Langevin equation \eqref{Langevin} around the hydrodynamic solution \eqref{density_1d}, i.e., writing $q=\rho+q_1$ and assuming that $q_1\ll \rho$, we find that the perturbation obeys a diffusion equation with a stochastic source 
\begin{equation}
\label{L_Lin}
     \partial_t q_1 - \partial_{xx} q_1=\partial_x \left[\sqrt{\sigma(\rho)} \,\xi(x,t)\right]
\end{equation}
Setting $q_1=\partial_x \psi$, we rewrite \eqref{L_Lin} as
\begin{equation}
\label{LL}
     \partial_t \psi - \partial_{xx} \psi=\sqrt{\sigma(\rho)} \,\xi(x,t)
\end{equation}
We must solve this equation subject to the initial condition $\psi(x, t=0)=0$ and the boundary condition 
$\partial_x \psi(x=0, t)=0$ which follows from the requirement $q(x=0, t)= \rho(x=0, t)=1$. The solution reads
\begin{eqnarray}
\psi(x,t) &=& \int_0^t dt^{\prime}\int_0^{\infty}dy\,\frac{\sqrt{\sigma(\rho)} \,\xi(y,t^{\prime})}
  {\sqrt{4 \pi (t-t^{\prime})}}\nonumber\\
&\times&\left[e^{-\frac{(x-y)^2}{4 (t-t^{\prime})}} + e^{-\frac{(x+y)^2}{4 (t-t^{\prime})}}\right]. 
\label{psi}
\end{eqnarray}
The total number of particles $N_+$ in the $x>0$ region is 
\begin{equation}
N_+ = \langle N_+\rangle + \int_0^\infty dx\,q_1(x,t)
\end{equation}
The variance is therefore 
\begin{eqnarray}
\langle N_+^2\rangle_c &=& \left\langle \int_0^{\infty} dx \int_0^{\infty} dy \,q_1(x,t) q_1(y,t) \right\rangle \nonumber \\
   &=& \left\langle  \int_0^{\infty} dx \int_0^{\infty} dy \, \partial_x \psi(x,t) \,\partial_{y} \psi(y,t) \right\rangle \nonumber \\
   &=& \langle \psi^2 (0,t)\rangle
   \label{var1}
\end{eqnarray}
Plugging \eqref{psi} into \eqref{var1} and using \eqref{noise} we obtain
\begin{equation}
\label{N+}
\langle N_+^2\rangle_c = \frac{1}{\pi}\int_0^t \frac{dt^\prime}{t-t^\prime}\int_0^\infty dy\,
\sigma[\rho(y,t^\prime)]\,e^{-\frac{y^2}{2(t-t^\prime)}}
\end{equation}
The total number of particles $N_-$ in the $x<0$ region is a random quantity which is independent on $N_+$ and identically distributed. Therefore for the total number of particles $N=N_-+N_+$ the variance is 
$\langle N^2\rangle_c = 2\langle N_+^2\rangle_c$. Using this together with \eqref{N+}  we obtain
\begin{equation}
\label{N_var}
\langle N^2\rangle_c = \frac{2}{\pi}\int_0^t \frac{d\tau}{\tau}\int_0^\infty dx\,
\sigma[\rho(x,t-\tau)]\,e^{-\frac{x^2}{2\tau}}
\end{equation}
Making the transformation $\tau=t T$ and $x=\sqrt{t}\,X$, and using $\sigma(\rho)=2\rho(1-\rho)$ together with Eq.~\eqref{density_1d}, we recover the time dependence $\langle N^2\rangle_c = V_1\sqrt{t}$  with amplitude
\begin{equation}
\label{var_2}
V_1 = \tfrac{4}{\pi}(U_1-U_2)
\end{equation}
where
\begin{equation*}
U_p=\int_0^1 \frac{dT}{T}\int_0^\infty dX\left[\text{erf}\left(\frac{X}{2\sqrt{1-T}}\right)\right]^p\,
e^{-\frac{X^2}{2T}}
\end{equation*}
Computing the first two integrals,
\begin{equation*}
U_1 = \sqrt{2\pi} - \sqrt{\pi}\,,\quad 
U_2 = \big(3-2\sqrt{2}\big)\sqrt{2\pi}\,, 
\end{equation*}
we simplify \eqref{var_2} to the announced result \eqref{var_1d}. 

\section{Summary}
\label{summary}

We studied the growth of the total number of particles in a symmetric exclusion process driven by a localized source. Specifically, we assumed that new particles are injected into a single lattice site, the origin, whenever the origin is empty. We showed that the average total number of particles entering an initially empty system exhibits a simple asymptotic growth \eqref{Nav}. In one and two dimensions, the leading asymptotic behaviors for the average total number of particles turn out to be universal (independent of the flux) and in both cases the asymptotic growth is slower than linear in time. In three and higher dimensions, the average total number of particles entering the system grows linearly with time, namely as $\Phi_d(F)\,t$. We derived a simple equation \eqref{JF_3} expressing the renormalized flux $\Phi_d(F)$ through the bare flux $F$.  

In one and two dimensions, the results are insensitive not merely to the flux $F$, but also to the detailed structure of the source (the injection may occur through a few sites; the number of such sites and the details of their location are irrelevant) and to the lattice structure (in two dimensions, there is no need to require that the lattice is a square grid). In three and higher dimensions, the results are sensitive to the aforementioned detailed properties. It seems that if the lattice is arbitrary, but new particles are still injected into a single lattice site, the expression \eqref{JF_3} for the renormalized flux remains valid, although one has to use a proper expression for the Watson integral depending on the underlying lattice; e.g., for the body centered cubic lattice (bcc) the Watson integral is 
\begin{eqnarray*}
W_\text{bcc} &=&\int_{-\pi}^\pi \frac{d^3 {\bf q}}{2(2\pi)^3}\,\frac{1}{1-\cos q_1\cos q_2\cos q_3}\\
&=& \frac{[\Gamma(\tfrac{1}{4})]^4}{(2\pi)^3} = 0.696601966\ldots
\end{eqnarray*}

We also discussed fluctuations of the total number of particles, especially the variance for which we gave  asymptotic growth laws \eqref{N_variance}. The arguments leading to \eqref{N_variance} are solid only in one dimension where we used fluctuating hydrodynamics and analytically established the asymptotic growth of the variance. In two dimensions, the result cited in Eq.~\eqref{N_variance} is just a guess with an unknown multiplicative factor $V_2$; even the functional form of the variance is conjectural. It would be interesting to compute the variance in two dimensions by employing fluctuating hydrodynamics or another continuum approach \cite{KM2011}. In three and higher dimensions, the prediction of Eq.~\eqref{N_variance} for the variance is sharp yet unproven; one would like to justify that prediction, or disprove it.

The one-dimensional case is particularly tractable, and one may be able to determine higher cumulants. To guess the outcome, we notice a similarity of our problem and the problem of the evolution of the SEP starting with a step-function initial condition, particularly $\rho(x, t=0)=1$ for $x<0$ and $\rho(x, t=0)=0$ for $x\geq 0$. In this latter problem all cumulants of the total current grow in a diffusive manner \cite{DG2009a,DG2009b}. The same should be valid for the cumulants of $N(t)$ in our problem. Since the numbers $N_+$ and $N_-$ of particles to the right and left of the origin are independent (in the simplest case of infinite flux) and identically distributed random variables, it suffices to consider $N_+(t)$. Similarly to the SEP with a step-function initial condition \cite{DG2009a} we anticipate that 
\begin{equation}
\label{Pn}
P(N_+,t)=\text{Prob}\left[\frac{N_+(t)}{\sqrt{t}}=n\right] \sim e^{- \sqrt{t}\, G(n)}
\end{equation}

The derivation of the large deviation function for the SEP with the step-function initial condition \cite{DG2009a} is  complicated and the results do not seem to admit a straightforward extension to the present case. For instance, the probability to have zero total current is $\propto e^{-\sqrt{t}}$ \cite{DG2009a}, which means that the large deviation function remains finite in the small current limit: $G_\text{step}(0)<\infty$. In our case, $N_+(t)=0$, or more generally that $N_+(t)$ is small, with probability $\propto e^{-t}$; this implies that the large deviation function diverges, $G(n)\sim n^{-1}$ as $n\to 0$. Both settings are still very close, so the methods of Ref.~\cite{DG2009a} could be applicable to the present case, at least in the situation when the flux is infinite. 

\bigskip
\noindent
I thank Kiron Mallick, Baruch Meerson, Gleb Oshanin, and Darko Stefanovic for useful correspondence.

\end{document}